\documentclass[RNAAS,twocolumn,times]{aastex62}

\newcommand\MbarI{\ensuremath{\overline{M}_I}}
\newcommand\MbarIacs{\ensuremath{\overline{M}_{814}}}

\newcommand\mbarIacs{\ensuremath{\overline{m}_{814}}}

\newcommand\vi{{\ifmmode{(V{-}I)}\else$(V{-}I)$\fi}}
\newcommand\gI{{\ifmmode{(g{-}I)}\else$(g{-}I)$\fi}}
\newcommand\gIacs{{\ifmmode{(g_{475}{-}I_{814})}\else$(g_{475}{-}I_{814})$\fi}}

\begin{document}

\title{Independent Analysis of the Distance to NGC1052-DF2}

\correspondingauthor{John Blakeslee, Michele Cantiello}
\email{jblakeslee@gemini.edu; cantiello@oa-abruzzo.inaf.it}

\author[0000-0002-5213-3548]{John P.\ Blakeslee}
\affil{Gemini Observatory, La Serena, Chile}

\author[0000-0003-2072-384X]{Michele Cantiello}
\affil{INAF Osservatorio Astronomico d’Abruzzo, via Maggini, snc, 64100 Italy}

\keywords{galaxies: distances and redshifts --- galaxies: individual (NGC\,1052-DF2) --- dark matter}

\section{Context} 

The finding by \citet[][hereafter~vD18a]{vdk18nat} that the dynamical mass of
the diffuse galaxy NGC1052-DF2, also called [KKS2000]04
\citep[see][hereafter T18]{trujillo18}, is consistent with its stellar mass,
``and leaves little room for a dark matter halo,'' has created a stir in the
extra\-galactic astronomy community.
The result remains controversial \citep[e.g.,][]{martin18}.
As stated by vD18a, this conclusion depends critically
on the adopted distance of 20~Mpc from association with NGC\,1052, which is consistent
with their surface brightness fluctuations (SBF) result $d=19.0\pm1.7$~Mpc
for NGC1052-DF2 itself. 

Following the publication of vD18a, we were contacted by several colleagues 
about the NGC1052-DF2 SBF \hbox{distance}. We noted some areas of concern: use of the
default linear drizzle interpolation kernel can bias the image power spectrum
measurement \citep{cantiello05,mei05iv}; information was lacking on the
residual variance correction; and the ACS/F814W SBF magnitude calibration
\citep[][hereafter B10]{blake10} was linearly extrapolated beyond the explored color range.

We therefore did a quick SBF analysis using our own well-documented procedures
\citep[see B10,][and references therein]{cantiello18} on this
galaxy of unusually low surface brightness and our ``best guess'' calibration.
Our result agreed with vD18a; thus, we saw no need to publish it. 
However, given the renewed interest sparked by the dueling works of
T18 and \citet[][hereafter vD18b]{vdk18apj}, we report a revised, more careful
analysis here.


\section{Distance Analysis} 

The sinc-like lanczos3
interpolation kernel is preferred for SBF analysis; drizzle-combined F606W/F814W
HST/ACS images of NGC1052-DF2 using this kernel were kindly provided by S.~Lim.
On inspection, the galaxy exhibits clear SBF, but not an abundantly resolved
stellar population.\footnote{Of course, our customary tool is SBF,
  and to a child with a hammer, everything looks like a nail.} 
We performed standard SBF and color measurements, described most recently in
\citet{cantiello18}, using a circular annular region of
$0\farcs4\hbox{--}12\farcs8$ centered on the photo-center of NGC\,1052-DF2.
Omitting any correction for background variance, we find
$\mbarIacs=29.57\pm0.13$ and
$V_{606}{-}I_{814}=0.39\pm0.02$, both AB~mag.

Our measurement is fainter than $\mbarIacs=29.45\pm0.10$~mag reported by vD18a, but
%
repeating our analysis using the default drizzle kernel yielded
$\mbarIacs=29.42$, very close to vD18a.
T18 adopted \mbarIacs\ from vD18a, but used for calibration $\MbarIacs\approx-1.4$,
significantly fainter than $-1.94$ (vD18a) or $-1.91$ (vD18b).
%
We note that the color of NGC1052-DF2, and the SED fitting
by T18, indicate its stellar population is consistent with
halo globular clusters (GCs), for which \citet{at94} determined 
$\MbarI=-2.02\pm0.04$, independent of mean color.
Revising this calibration with an updated RR~Lyrae magnitude-metallicity relation
\citep{clementini03} and the LMC distance modulus of $18.49\pm0.05$~mag from DEBs
\citep{pietrzynski13} gives $\MbarI=-2.22\pm0.06$.  Converting from Cousins~$I$
to F814W \citep{saha11} with the latest ACS zeropoints then gives
$\MbarIacs=-1.78\pm0.12$ ABmag.


Our $V_{606}{-}I_{814}$ measurement is consistent with vD18a; applying their color
transformation gives $\gIacs=0.82\pm0.04$, consistent with $\gIacs=0.85\pm0.02$
from T18. Using the latter value with linear extrapolation of the B10 
calibration gives $\MbarIacs=-1.81$, but we agree with T18 this is
risky; the models plotted by T18 diverge towards $\MbarIacs{<\,}-2$ at these
colors.  However, the SPoT models used by B10 agree well with the empirical
calibration and suggest a natural extension; at $\gIacs=0.85$, they
predict $\MbarIacs=-1.75\pm0.05$, in excellent agreement with the GC-based
calibration above.

Using $\mbarIacs=29.57\pm0.13$ and our GC-based $\MbarIacs$
calibration, the NGC1052-DF2 distance is $18.6\pm1.9$~Mpc.  For the sake of
comparison to vD18a, we have not included any correction for background
contamination from faint sources.  Our standard procedure applied
blindly to this diffuse galaxy indicated a correction of ${\sim\,}0.4$~mag,
resulting in the preliminary value of ${\sim\,}22$ Mpc
quoted on PvD's webpage.\footnote{https://www.pietervandokkum.com/ngc1052-df2} 
On closer inspection, the correction is more likely $0.2\pm0.1$~mag, giving
a final distance $d=20.4\pm2.0$~Mpc.

\section{Commentary} 

Our \mbarIacs\ is 0.12~mag fainter than vD18a's, while our adopted
\MbarIacs\ calibration is 0.13~mag fainter than vD18b's, giving a distance, prior to
background variance correction, indistinguishable from vD18b's. The difference in
the measured SBF magnitudes is consistent with the shift we find when analyzing
images drizzled with a linear kernel rather than the lanczos3 kernel.
Since all the vD18b SBF measurements use the linear kernel, their distance
ladder is internally consistent.  As a further check, we measured the difference in
SBF magnitude between NGC1052-DF2 and M96-DF11; our result agreed with theirs 
within errors.  Correcting for background variance increases our NGC1052-DF2
distance by 10\% to ${\sim\,}20$~Mpc.

The GCs remain intriguing: if they're ``normal,'' NGC1052-DF2 is at
${\sim\,}$12~Mpc with substantial dark matter; if the distance is 
20~Mpc, the GCs are bigger and brighter than normal.  If we had a bias, it was that
any galaxy rich in GCs should be dominated by dark matter, as one of us proposed two
decades ago that the number of GCs scales with galaxy halo mass.
However, our distance analysis supports the conclusion that the stars can
account for the entire dynamical mass, although more kinematic data are needed.

While we think the evidence is strong that $d>16$~Mpc (at $2\sigma$), the
result is not quite definitive. The SBF method is not well-tested at these colors
and low stellar densities, although the vD18b team has made impressive progress. 
Extremely deep \textit{Hubble} data would resolve the issue by providing a definitive
detection of the TRGB.

\acknowledgments
We thank Eric Peng, Mike Beasley, \& Pieter van~Dokkum for helpful
conversations and Sungsoon~Lim for drizzling.

\end{document}